\documentclass[12pt]{article}
\usepackage{epsfig}
\usepackage{amsmath}
\usepackage{hhline}
\usepackage{amssymb}
\usepackage{times}
\usepackage{cite}
\usepackage{color}
\usepackage{lineno}
\usepackage{rotating}

\newlength{\dinwidth}
\newlength{\dinmargin}
\begin{document}  
\newcommand{\pom}{{I\!\!P}}
\newcommand{\reg}{{I\!\!R}}
\newcommand{\slowpi}{\pi_{\mathit{slow}}}
\newcommand{\fiidiii}{F_2^{D(3)}}
\newcommand{\fiidiiiarg}{\fiidiii\,(\beta,\,Q^2,\,x)}
\newcommand{\n}{1.19\pm 0.06 (stat.) \pm0.07 (syst.)}
\newcommand{\nz}{1.30\pm 0.08 (stat.)^{+0.08}_{-0.14} (syst.)}
\newcommand{\fiidiiiful}{F_2^{D(4)}\,(\beta,\,Q^2,\,x,\,t)}
\newcommand{\fiipom}{\tilde F_2^D}
\newcommand{\ALPHA}{1.10\pm0.03 (stat.) \pm0.04 (syst.)}
\newcommand{\ALPHAZ}{1.15\pm0.04 (stat.)^{+0.04}_{-0.07} (syst.)}
\newcommand{\fiipomarg}{\fiipom\,(\beta,\,Q^2)}
\newcommand{\pomflux}{f_{\pom / p}}
\newcommand{\nxpom}{1.19\pm 0.06 (stat.) \pm0.07 (syst.)}
\newcommand {\gapprox}
   {\raisebox{-0.7ex}{$\stackrel {\textstyle>}{\sim}$}}
\newcommand {\lapprox}
   {\raisebox{-0.7ex}{$\stackrel {\textstyle<}{\sim}$}}
\def\gsim{\,\lower.25ex\hbox{$\scriptstyle\sim$}\kern-1.30ex%
\raise 0.55ex\hbox{$\scriptstyle >$}\,}
\def\lsim{\,\lower.25ex\hbox{$\scriptstyle\sim$}\kern-1.30ex%
\raise 0.55ex\hbox{$\scriptstyle <$}\,}
\newcommand{\pomfluxarg}{f_{\pom / p}\,(x_\pom)}
\newcommand{\dsf}{\mbox{$F_2^{D(3)}$}}
\newcommand{\dsfva}{\mbox{$F_2^{D(3)}(\beta,Q^2,x_{I\!\!P})$}}
\newcommand{\dsfvb}{\mbox{$F_2^{D(3)}(\beta,Q^2,x)$}}
\newcommand{\dsfpom}{$F_2^{I\!\!P}$}
\newcommand{\gap}{\stackrel{>}{\sim}}
\newcommand{\lap}{\stackrel{<}{\sim}}
\newcommand{\fem}{$F_2^{em}$}
\newcommand{\tsnmp}{$\tilde{\sigma}_{NC}(e^{\mp})$}
\newcommand{\tsnm}{$\tilde{\sigma}_{NC}(e^-)$}
\newcommand{\tsnp}{$\tilde{\sigma}_{NC}(e^+)$}
\newcommand{\st}{$\star$}
\newcommand{\sst}{$\star \star$}
\newcommand{\ssst}{$\star \star \star$}
\newcommand{\sssst}{$\star \star \star \star$}
\newcommand{\tw}{\theta_W}
\newcommand{\sw}{\sin{\theta_W}}
\newcommand{\cw}{\cos{\theta_W}}
\newcommand{\sww}{\sin^2{\theta_W}}
\newcommand{\cww}{\cos^2{\theta_W}}
\newcommand{\trm}{m_{\perp}}
\newcommand{\trp}{p_{\perp}}
\newcommand{\trmm}{m_{\perp}^2}
\newcommand{\trpp}{p_{\perp}^2}
\newcommand{\alp}{\alpha_s}

\newcommand{\alps}{\alpha_s}
\newcommand{\sqrts}{$\sqrt{s}$}
\newcommand{\LO}{$O(\alpha_s^0)$}
\newcommand{\Oa}{$O(\alpha_s)$}
\newcommand{\Oaa}{$O(\alpha_s^2)$}
\newcommand{\PT}{p_{\perp}}
\newcommand{\JPSI}{J/\psi}
\newcommand{\sh}{\hat{s}}
\newcommand{\uh}{\hat{u}}
\newcommand{\MP}{m_{J/\psi}}
\newcommand{\PO}{I\!\!P}
\newcommand{\xbj}{x}
\newcommand{\xpom}{x_{\PO}}
\newcommand{\ttbs}{\char'134}
\newcommand{\xpomlo}{3\times10^{-4}}  
\newcommand{\xpomup}{0.05}  
\newcommand{\dgr}{^\circ}
\newcommand{\pbarnt}{\,\mbox{{\rm pb$^{-1}$}}}
\newcommand{\gev}{\,\mbox{GeV}}
\newcommand{\WBoson}{\mbox{$W$}}
\newcommand{\fbarn}{\,\mbox{{\rm fb}}}
\newcommand{\fbarnt}{\,\mbox{{\rm fb$^{-1}$}}}
\newcommand{\unit}[1]{\ {\rm#1}} 
%
%
\newcommand{\ye}{$y$}
\newcommand{\qsq}{\ensuremath{Q^2} }
\newcommand{\gevsq}{\ensuremath{\mathrm{GeV}^2} }
\newcommand{\et}{\ensuremath{E_t^*} }
\newcommand{\rap}{\ensuremath{\eta^*} }
\newcommand{\gp}{\ensuremath{\gamma^*}p }
\newcommand{\dsiget}{\ensuremath{{\rm d}\sigma_{ep}/{\rm d}E_t^*} }
\newcommand{\dsigrap}{\ensuremath{{\rm d}\sigma_{ep}/{\rm d}\eta^*} }
\def\Journal#1#2#3#4{{#1} {\bf #2} (#3) #4}
\def\NCA{\em Nuovo Cimento}
\def\NIM{\em Nucl. Instrum. Methods}
\def\NIMA{{\em Nucl. Instrum. Methods} {\bf A}}
\def\NPB{{\em Nucl. Phys.}   {\bf B}}
\def\PLB{{\em Phys. Lett.}   {\bf B}}
\def\PRL{\em Phys. Rev. Lett.}
\def\PRD{{\em Phys. Rev.}    {\bf D}}
\def\ZPC{{\em Z. Phys.}      {\bf C}}
\def\EJC{{\em Eur. Phys. J.} {\bf C}}
\def\CPC{\em Comp. Phys. Commun.}

%
\newcommand{\Dstar}{\particle{D^{*}}}
\newcommand{\Dstarplus}{\particle{D^{*+}}}
\newcommand{\Dsubs}{\particle{D_s}}
\newcommand{\Dsubsplus}{\particle{D^{+}_s}}
\newcommand{\Dzero}{\particle{D^{0}}}
\newcommand{\Dplus}{\particle{D}^+}
\newcommand{\BR}{{\cal B}}
\newcommand{\GeV}{\mathrm{GeV}}
\newcommand{\mev}{\mathrm{MeV}}
\newcommand{\dst}{$D^{*+}$}
\newcommand{\dc}{$D^+$}
\newcommand{\dn}{$D^0$}
\newcommand{\ds}{$D^+_s$}
\newcommand{\dstdec}{$D^{*+} \rightarrow D^0 \pi^+_s \rightarrow (K^- \pi^+) \pi^+_s$}
\newcommand{\dcdec}{$D^+ \rightarrow K^- \pi^+ \pi^+$}
\newcommand{\dndec}{$D^0 \rightarrow K^- \pi^+$}
\newcommand{\dsdec}{$D^+_s \rightarrow \phi \pi^+ \rightarrow (K^+ K^-) \pi^+$}
\newcommand{\particle}[1]{#1}
\newcommand{\cbar}{\particle{\bar{c}}}
\newcommand{\bbar}{\particle{\bar{b}}}
\newcommand{\ccbar}{c\cbar}
\newcommand{\bbbar}{b\bbar}
\newcommand{\pb}{\rm pb}
\newcommand{\icaption}[1]{\caption{\it #1}}

%

%

\begin{titlepage}

\noindent

\begin{figure}[!t]
DESY 04--156 \hfill ISSN 0418--9833\\
August 2004
\end{figure}
\bigskip

\vspace*{2cm}

\begin{center}
\begin{Large}

{\bf Inclusive Production of {\boldmath $D^+, D^0,  D_s^+$} and {\boldmath $D^{*+}$} Mesons \\
in Deep Inelastic Scattering at HERA}

\vspace{2cm}

H1 Collaboration

\end{Large}
\end{center}

\vspace{2cm}

\begin{abstract}

Inclusive production cross sections are measured in deep inelastic
scattering at HERA for 
meson states composed of a charm quark and a light antiquark or the 
charge conjugate.
The measurements cover the kinematic region of 
photon virtuality  $2 < Q^2 < 100 \ \gev^2$,
inelasticity $0.05 < y < 0.7$,
$D$ meson  transverse momenta $p_t(D) \ge 2.5 \ \gev$ 
and  pseudorapidity $|\eta(D)| \le 1.5$.
The identification of the  $D$-meson decays and the reduction of 
the combinatorial background profit from the reconstruction
of displaced secondary vertices by means of the H1 silicon
vertex detector.
The production of charmed mesons
containing the light quarks $u,d$ and $s$ is found to be compatible with a
description in which the hard scattering is followed by a factorisable 
and universal hadronisation process.

\end{abstract}

\vspace{1.5cm}

\begin{center}
To be submitted to \EJC
\end{center}

\end{titlepage}

%
%

\begin{flushleft}

A.~Aktas$^{10}$,               
V.~Andreev$^{26}$,             
T.~Anthonis$^{4}$,             
A.~Asmone$^{33}$,              
A.~Babaev$^{25}$,              
S.~Backovic$^{37}$,            
J.~B\"ahr$^{37}$,              
P.~Baranov$^{26}$,             
E.~Barrelet$^{30}$,            
W.~Bartel$^{10}$
% ...

\bigskip{\it
% ...
$ ^{4}$ Inter-University Institute for High Energies ULB-VUB, Brussels;
          Universiteit Antwerpen, Antwerpen; Belgium$^{ c}$ \\
% ...
$ ^{10}$ DESY, Hamburg, Germany \\
% ...
$ ^{25}$ Institute for Theoretical and Experimental Physics,
          Moscow, Russia$^{ l}$ \\
 $ ^{26}$ Lebedev Physical Institute, Moscow, Russia$^{ e}$ \\
% ...
$ ^{30}$ LPNHE, Universit\'{e}s Paris VI and VII, IN2P3-CNRS,
          Paris, France \\
% ...
$ ^{33}$ Dipartimento di Fisica Universit\`a di Roma Tre
          and INFN Roma~3, Roma, Italy \\
% ...
$ ^{37}$ DESY, Zeuthen, Germany \\
% ...
$ ^c$ Supported by FNRS-FWO-Vlaanderen, IISN-IIKW and IWT
      and  by Interuniversity Attraction Poles Programme,
      Belgian Science Policy \\
% ...
$ ^e$ Supported by the Deutsche Forschungsgemeinschaft \\
% ...
$ ^l$ Partially Supported by Russian Foundation
      for Basic Research, grant    no. 00-15-96584 \\
}

\end{flushleft}

\newpage

\section{Introduction}
\noindent

\begin{figure}[hbt]
\begin{center}
    \epsfig{file=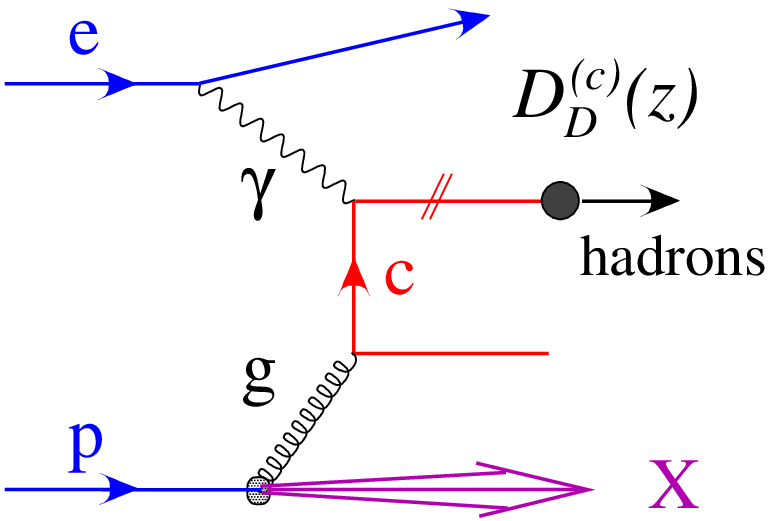,width=9cm}
\icaption{Diagram for the production of $c$-quarks in $ep$ collisions 
 via the photon-gluon fusion process. 
The hadronisation into a charmed meson $D$ is described by the fragmentation function
$D_D^{(c)}(z)$ (cf. Eq.~\ref{eq:f3}). }
  \label{fig:bgf}
\end{center}
\end{figure}

The production of heavy quarks in deep inelastic 
positron-proton interactions  at HERA proceeds 
almost exclusively via photon-gluon fusion, where 
a photon coupling to the incoming positron interacts 
with a gluon in the proton to form a charm-anticharm pair 
(Fig.~\ref{fig:bgf}).
Measurements of differential charm production cross 
sections ~\cite{Aid:1996hj,Adloff:1998vb,Adloff:2001zj,
Breitweg:1999ad,Chekanov:2003rb}
are well reproduced by a description based on perturbative 
QCD (pQCD).
In this framework cross sections d$\sigma$ for $D$ mesons
(bound  $c\bar{q}$ states of charm quarks with light antiquarks) can be
computed\cite{Ellis:1988sb,Smith:1991pw,Laenen:1992zk,Frixione:1993dg,
Frixione:1994nb,Frixione:nn}\ by convoluting the parton 
level hard scattering cross section
d$\hat{\sigma}$ with a fragmentation function $D_D^{(c)}(z)$:

\begin{eqnarray}
  {\rm d}\sigma(p) & = & 
   \int {\rm d}z \ \  D_D^{(c)}(z) 
   \cdot {\rm d}\hat{\sigma}(p/z)\ \ \ \ .  \label{eq:f1}
\end{eqnarray}
\noindent
The cross section d$\hat{\sigma}$ describes the hard scattering
process $(ep \to e' c X')$ and the perturbative 
evolution from the $c$-quark production scale down to a scale of order 
of the heavy quark mass $m_c$\cite{Nason:1999zj}.
The non-perturbative part of the hadronisation, $D_D^{(c)}(z)$,
accounts for the transition of an on-shell quark $c$ of momentum $p/z$
into a hadron $D$  carrying a fraction $z$ of the quark momentum. 
%

This factorisation ansatz can be shown to hold in the asymptotic limit
of large production scales~\cite{collins}.
It assumes that the non-perturbative component
$D_D^{(c)}(z)$ is independent both of the hard scattering process, e.g.
$ee-, ep-$ or $ pp-$scattering, and of
the scale at which the charm quark is produced.
These properties are collectively referred to as 
``universality''\cite{Frixione:nn,Mele:1990cw,Nason:1993xx,Nason:1999zj}.
It has been argued 
\cite{Nason:1999zj,n-mellin}
that the assumptions of universality may not hold and that in fact
different production processes may be sensitive to different aspects of
fragmentation. 

Conventionally the 
shape of $D_D^{(c)}(z)$ is taken to be identical for all primary produced 
$D$-mesons, and its normalisation  accounts for the abundance of the
respective meson $D$.
A well-known parametrisation of this shape 
is the Peterson function\cite{Peterson:1982ak} with a suitably chosen
normalisation factor $n_D$

\begin{eqnarray}
   D_D^{(c)}(z) & = & \frac{n_D}{ z [ 1-1/z-\epsilon_{c}/(1-z)]^2}\ \ \  \ .   
   \label{eq:f3}
\end{eqnarray}

\noindent
The so-called ``Peterson parameter'' $\epsilon_{c}$ measures 
the hardness of the hadronisation. In practice, it is extracted from 
charm spectra measured predominantly at 
$e^+e^-$ colliders\cite{Nason:1999zj}.

\noindent
In this paper the universality ansatz is explored in 
deep inelastic $ep$-scattering at HERA.
For this purpose production cross sections are measured for the charmed meson state 
 $D^+, D^0,  D_s^+$ and $D^{*+}$,
containing the light quarks $u, d$ and $s$.
The final state requirements differ only in the light quark content
or the spin of the charmed meson.
For light quark fragmentation the occurrence of
different quark flavours
in the fragmentation process
can be described by isospin relations 
and probabilistic quark selection factors.
While corresponding charm quark fragmentation
results are available from
other scattering 
processes\cite{Nason:1999zj,Albrecht:1991ss,Akers:1994jc,Buskulic:1995gp},
it is interesting to verify whether the relations established for
light quarks can be extended 
to mesons containing heavy quarks produced in $ep$ scattering at HERA.

In this analysis the signals of $D$ mesons are enhanced with
respect to background processes by exploiting their lifetimes.
The displaced secondary decay vertices 
of the $D$-mesons are reconstructed from the daughter particle tracks,
which are measured with high precision in the H1 central silicon tracker.  
A common detection method can thus be applied to all four charmed meson states.
%
%

\section{Experimental Aspects and Data Analysis}


The data were collected in 1999 and 2000 with the H1 detector at HERA and
correspond  to an integrated luminosity of $47.8 \pm
0.7~\pb^{-1} $ of $e^+p$ interactions  at a
centre of mass energy $\sqrt{s} = 319~\GeV$.  
The analysis covers the kinematic region of photon virtuality 
$2 < Q^2 < 100 \ \gev^2$, inelasticity $0.05 < y < 0.7$
and  $D$ meson transverse momentum and 
pseudorapidity\footnote{In the 
following, the coordinate system used has its origin
at the nominal $e^+p$ interaction point and its
$z$-axis in the outgoing proton direction. Polar angles, $\theta$, 
are measured with respect to the $z$ direction. The pseudorapidity is
defined as $\eta = - \ln \tan(\theta /2)$.}
$p_t(D) \ge 2.5 \ \gev$ and
$|\eta(D)| \le 1.5$.
The charmed mesons are detected through their decay products 
in the decay channels\footnote{The
charge conjugate states are always implicitly
included.} \dcdec, \dndec, \dsdec\ and 
$D^{*+}(2010) \rightarrow D^0 \pi^+ \rightarrow (K^- \pi^+) \pi^+$.

\subsection{Detector and Simulation }

The H1 detector and its trigger capabilities are
described in detail elsewhere~\cite{Abt:1996hi}.
Charged particles are  measured by a set of central tracking chambers
including two cylindrical jet drift chambers (CJC) \cite{Burger:eb,Abt:1994dn}, 
mounted concentrically around the
beam-line inside a homogeneous magnetic field of 1.15~T.
They provide particle charge and transverse momentum measurements in the polar angle range
$20^{\circ}<\theta<160^{\circ}$.
Further drift chambers provide accurate measurements of the $z$-coordinates.

These central tracking chambers are complemented by the 
central silicon tracker (CST)~\cite{Pitzl:2000wz},
which consists of two cylindrical layers of 
silicon strip detectors with active lengths of $35.6$~cm
surrounding the beam pipe  
at radii of $57.5$~mm and $97 $~mm from the beam axis. 
The double-sided strip detectors 
measure $r$-$\phi$ and $z$ coordinates 
with resolutions of 12~$\mu$m and 25~$\mu$m, respectively.
Average hit efficiencies are 97 (92)\% for $r$-$\phi$ ($z$) strips.
In this analysis the measurements of the $z$-coordinates are not used for the
secondary vertex determination.
When extrapolated to the interaction region, the 
coordinate perpendicular to a track in the transverse
plane has a resolution of
$ 33\;\mu\mbox{m} \oplus 90 \;\mu\mbox{m} /p_t [\mbox{GeV}]$
for tracks with CST hits in both layers.
The first term represents the intrinsic resolution
and includes the uncertainty on the CST alignment.
The second term accounts for the contribution
from multiple scattering.

One double layer of cylindrical multi-wire proportional 
chambers (MWPC\cite{Muller:jk}) with pad readout
for triggering purposes is positioned between the CST and the CJC, and
another between the two jet chambers.
The backward region
of H1 is equipped with a lead/scintillating fibre
``Spaghetti'' Calorimeter (SpaCal)~\cite{Appuhn:1996na},
which is optimized for the detection of the scattered positron in the 
DIS kinematic range considered here. 
It consists of an electromagnetic and a more coarsely 
segmented hadronic section.
An eight-layer drift chamber (BDC~\cite{bdc}),
mounted in front of the SpaCal, is used to reject neutral particle background.
The $ep$ luminosity is determined by measuring the  QED
bremsstrahlung $(ep \to ep\gamma)$ event rate by tagging the
photon in a photon detector located at $z = -103$\ m.

Corrections for detector effects are evaluated
using detailed Monte Carlo simulations. 
Charm and beauty electroproduction events are generated according to 
Leading Order (LO) QCD matrix elements with the
AROMA 2.2~\cite{Ingelman:1996mv} program and are combined with parton showering
in the JETSET\cite{Sjostrand:1993yb} program.  
The beauty cross section is enhanced by a factor 4.3 in accordance
with the measurement of ~\cite{h1-bxsec-dis}.
The hadronisation step proceeds according to the
Lund string model for light quarks and the Peterson model for heavy
quarks. The simulation is adjusted to reproduce the world average
results on the relative abundances of charmed 
mesons~ \cite{Gladilin:1999pj}.
The generated events are then processed by the H1 detector 
simulation program and are subjected to the same reconstruction 
and analysis chain as the real data.


\subsection{Selection of DIS Events}

The analysis is based on events which contain a scattered positron
detected in the SpaCal backward calorimeter.
Scattered positrons are identified as clusters in the SpaCal 
with energies $E_{e^{\prime}}>8$ GeV, with cluster radii less than 
3.5~cm, consistent with
an electromagnetic energy deposition.
The cluster centres must be
spatially associated with a charged track candidate in the BDC
within the polar angular region $153^{\circ}<\theta<177.8^{\circ}$. 
The events are triggered by the identification of a
SpaCal cluster in coincidence with a central charged track signal
from the MWPC and central drift chamber triggers.

\noindent
At fixed $\sqrt{s}$
the kinematics of the scattering process are determined by 
two of the Lorentz invariant variables $Q^2$, $y$ and the 
Bjorken scaling variable $x$. 
These variables are reconstructed using the scattered 
positron according to 

\begin{equation}
Q^2  = 4 E_e E_{e^{\prime}} \cos ^2 \left( \frac{\theta_{e^{\prime}}}{2} \right)  ,
  {\rm \quad  } 
y = 1 - \frac{E_{e^{\prime}}}{E_e} \sin ^2
\left( \frac{\theta_{e^{\prime}}}{2} \right),  \;\; \quad
x = \frac{Q^2}{ys},
\label{eq:kine}
\end{equation}
with $E_e$ and $E_{e^{\prime}}$ denoting the energies of the incoming and scattered positron,
respectively, and $\theta_{e^{\prime}}$ the polar angle of the
scattered positron.


\subsection{Decay Vertex Reconstruction}

The lifetimes  of between 0.4 and 1 ps of the $D$ mesons lead to a
spatial separation between their production vertex,
assumed to be the point of the primary $ep$ interaction,
and the decay vertex, referred to as the secondary vertex.
This separation is expressed here in terms of the radial decay length
$l$ and its error $\sigma_l$.
The decay length measures the distance between these two vertices in the 
$r$-$\phi$ plane, which is typically a few hundred $\mu$m.
The significance $S_l = l/\sigma_l$ is a 
powerful discriminator to identify long-lived hadrons\cite{Gassner:2002ji}.

Tracks are first reconstructed in the CJC and are then extrapolated
into the CST, where  the closest hit from strips measuring 
$r$-$\phi$ in each layer is associated with each track 
if it lies within $5\sigma$ 
of the extrapolated CJC track. 
Then the tracks are refitted using the additional information of the CST hits.
Two or three such CST-improved tracks are used to form a $D$ meson
candidate by fitting them to a common secondary vertex in the
$r$-$\phi$ plane, while all other tracks in the event,
CST-improved or not, are used for the determination of the primary vertex.

For the primary vertex fit an iterative procedure is adopted.
The track with the 
largest contribution to the $\chi^2$
is removed and the fit is repeated until the 
$\chi^2$ contribution of each track assigned to the 
primary vertex is less than three.
The precisely known position and profile of the elliptical 
interaction region in the $r$-$\phi$ plane (``beam spot'')
enter the primary vertex fit as an additional measurement 
of the vertex coordinates with uncertainties given by the 
width of the interaction region.

The position of the beam spot is determined 
run-wise or for a large number of events inside a run
by averaging the reconstructed $ep$ interaction points.
The profile of the beam spot is assumed to
be Gaussian and stable throughout the data-taking period.
Its width is determined from a sample of events where the primary vertex
is accurately measured from high momentum central tracks, which 
are largely unaffected by multiple scattering. 
Widths of $145\ \mu$m and $25\ \mu$m are found for 
the horizontal and vertical directions, respectively.
%
The uncertainty of the primary vertex determination varies from event
to event with an average $69\ \mu$m in the horizontal direction.
The uncertainty in the vertical direction is essentially given by the
$25\ \mu$m vertical width of the beam spot.

In the fit for the secondary vertex the momentum of
the $D$ meson candidate is constrained to be parallel (or
anti-parallel) to the vector connecting the primary and 
secondary vertices.
This provides a powerful constraint to reject random track combinations.
The primary vertex position and its error are used as external
measurements in the fit.
The uncertainty on the decay length is dominated by the resolution of
the secondary vertex reconstruction.

$D$ meson decay candidates are selected by applying requirements 
on the following variables:
the minimal transverse momentum $p_t$
of daughter tracks, 
a signed impact parameter\footnote{The
impact parameter $d$
is the distance of closest approach of a track to the primary vertex.
It is positive if the angle between the momentum vector
of the $D$ meson candidate and the vector pointing from the primary
vertex to the point of closest approach of the track
is less than $90^\circ$.}
$d$ of significance $S_d = d/\sigma_d$ for at least  two tracks, 
the secondary vertex fit probability  ${\cal P}_{\it VF}$,  
the error on the decay length measurement $\sigma_l$
and the vertex separation significance $S_l = l/\sigma_l$.
The detailed cuts, chosen to optimise the signal significance, 
are listed in Table~\ref{tab:selection}.
At most one missing CST hit is allowed for each $D$
meson candidate. The CST geometry confines 
the $D$ mesons to lie in a polar angular range of 
typically $25^\circ < \theta < 155 ^\circ$.
%
%

\begin{figure}[htb]
  \begin{center}
    \epsfig{file=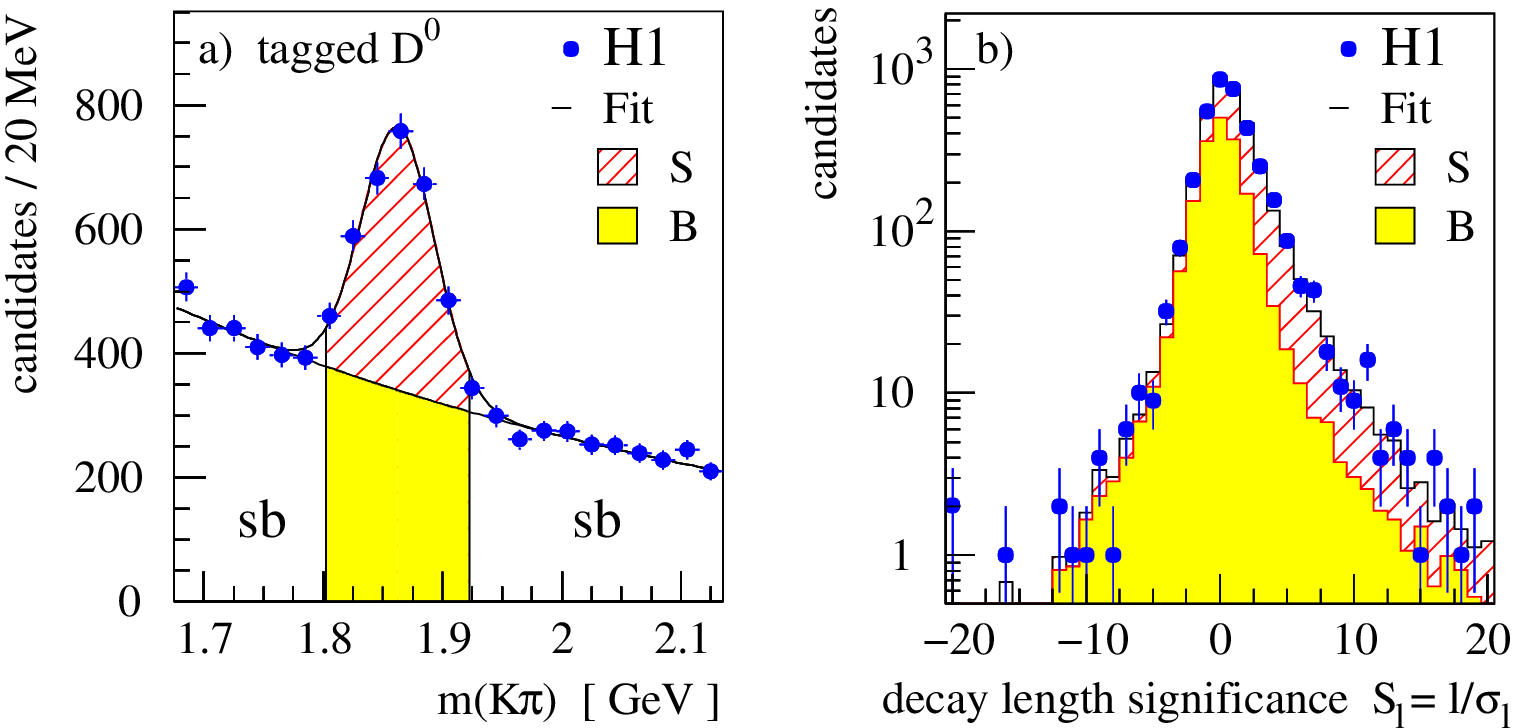,width=16cm}
\icaption{ 
 a) Invariant
 mass distribution $m(K\pi)$  
and  b) signed decay length significance distribution $S_l$
measured for tagged $\Dzero$
 mesons in data (solid points).
 Superimposed is the fitted decomposition into signal
($S$, hatched) and background ($B$, shaded) contributions (see text).
 The background in b) is determined from the $m(K\pi)$ sideband
regions $(sb)$ as indicated in a). }
  \label{fig:eff-sl}
  \end{center}
\end{figure}

%
To test the understanding of the CST,
its response to data has been compared with simulations for a sample
of ``tagged'' $\Dzero$  mesons (Fig.~\ref{fig:eff-sl}a). 
These $\Dzero$ mesons originate from 
a \dstdec\ decay chain and are selected by a cut about the nominal value
of $\Delta m = m(K \pi \pi_s) - m(K \pi)$  corresponding
to a window of three times the resolution.
In this channel a good signal purity can be achieved without a
lifetime cut, thus allowing a  determination of the efficiency
of the CST reconstruction.
Fig.~\ref{fig:eff-sl}b shows the measured decay length significance
distribution  $S_l$ of tagged $D^0$ candidates, together with 
its fitted decomposition into a signal and a background  contribution.
The functional form of the signal
distribution is taken from  the simulation,
whereas the background shape is extracted from the 
sideband regions on both sides of the nominal
\dn\ mass in the $m(K\pi)$ spectrum of the data, as indicated in 
Fig.~\ref{fig:eff-sl}a.
In the fit of the $S_l$ distribution 
only the normalisations of the signal and of the
background are left as free parameters.
The $\chi^2/{\it ndf} = 34/32$ indicates that the MC simulation 
describes the signal shape very well. 
Furthermore, the number of $\Dzero$ candidates extracted by means of the
$S_l$ fit is found to be fully consistent with
the number determined from the fit shown in 
Fig.~\ref{fig:eff-sl}a) 
to the invariant mass distribution $m(K\pi)$. 

\begin{figure}[hhh]
  \begin{center}
    \epsfig{file=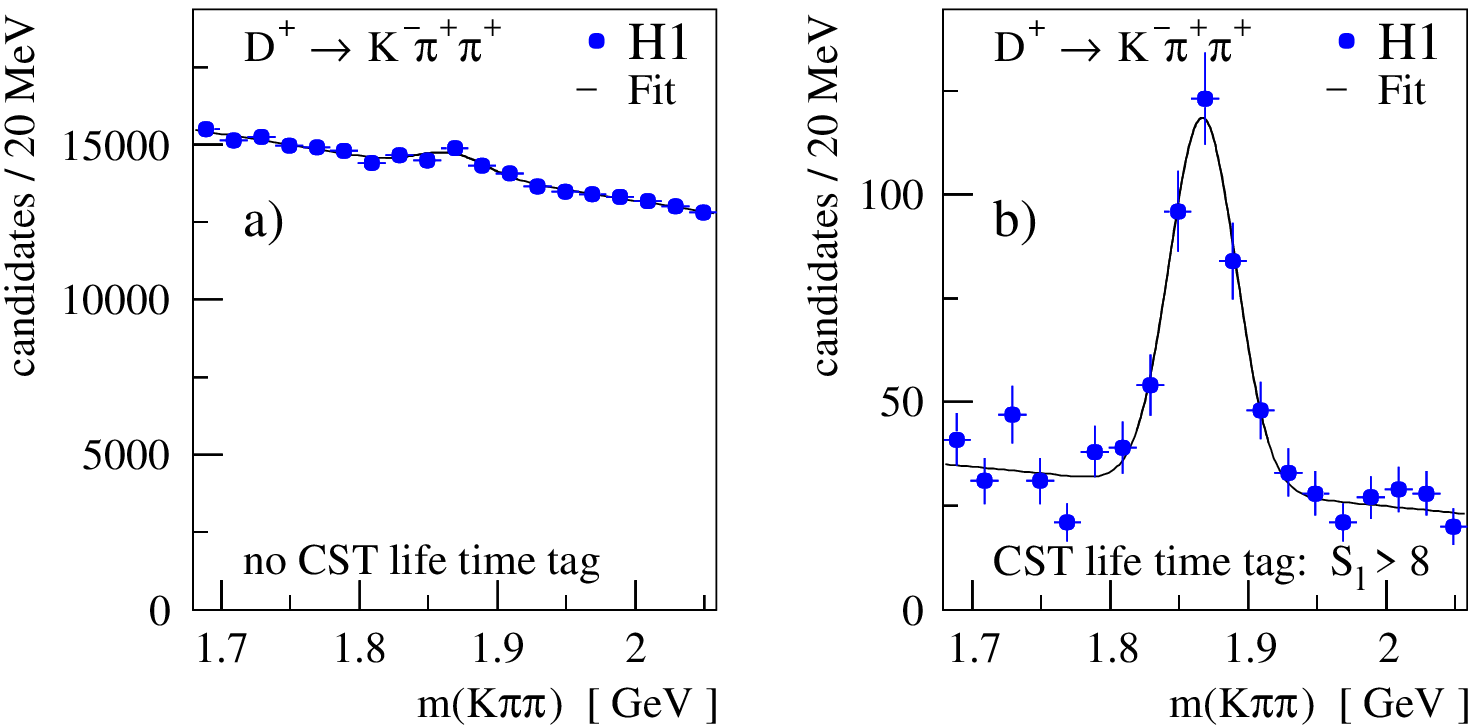,width=16. cm}
\icaption{ 
 Invariant mass distributions $m(K \pi \pi)$ for
 $D^+ \rightarrow K^- \pi^+ \pi^+$ decay candidates
(a) before and (b) after a cut on the decay length significance $S_l> 8$.}
  \label{fig:phenix}
  \end{center}
\end{figure}

To illustrate the purity that can be reached,
the invariant mass distributions for 
$\Dplus \to K^+ \pi^- \pi^+$ candidate events
are compared  before and 
after a stringent cut $S_l>8$ in Fig.~\ref{fig:phenix}.
The signal-to-background ratio improves by a factor 50 while
20\% of the signal is retained.
For the cross section measurements described below a less stringent
cut $S_l>5$ is chosen to increase the efficiency, albeit with a reduced purity.


\subsection{Signal Determination}

\begin{figure}[tb] 
  \begin{center}
    \epsfig{file=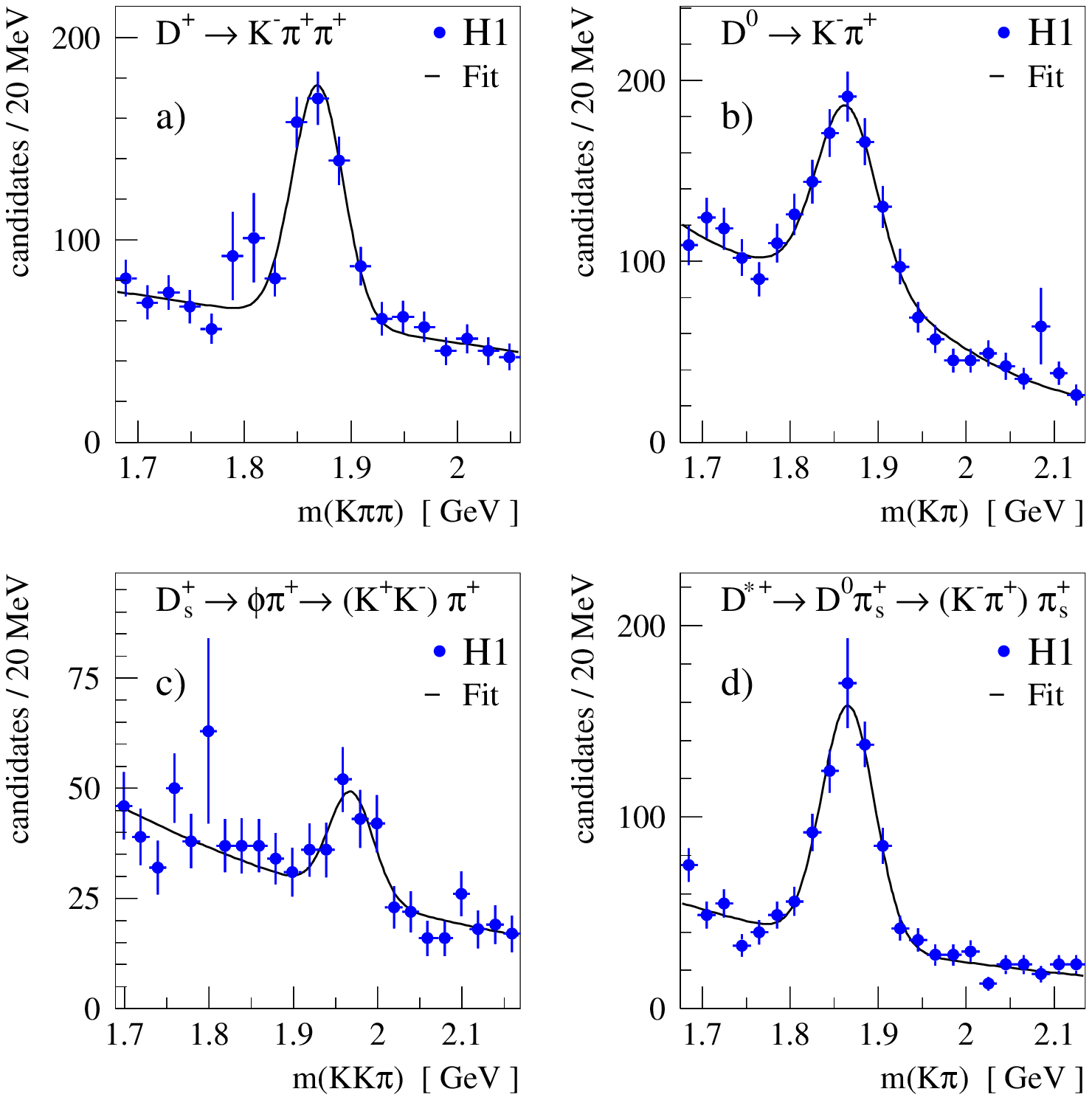}
  \end{center}
  \icaption{ 
  Invariant mass distributions for the four $D$ meson candidate decays
    that are used to determine the number of signal events:
a) $\Dplus \to K^- \pi^+ \pi^+$, 
b) $\Dzero \to K^- \pi^+$, 
c) $\Dsubsplus \to \phi \pi^+ \to K^+ K^- \pi^+$  and
d) $\Dstarplus \to \Dzero \pi^+_s \to K^- \pi^+ \pi^+_s$ 
  ( the invariant mass $m(K \pi)$ is shown after the $\Delta m$ cut).
The curves show the fits in which the signals are described by a
Gaussian and the backgrounds are parametrised as discussed in the text.
}
  \label{fig:dall-cst}
\end{figure}

The reconstruction of $D$ mesons uses all CST-improved 
charged particle tracks.
No explicit particle identification is applied, and
each track is assumed to be either a kaon or a pion as appropriate.
The tracks of suitable charge are required to pass the 
transverse momentum criteria specified 
in Table~\ref{tab:selection}.
They are then fitted to a common secondary vertex, which must also
fulfil the requirements listed in Table~\ref{tab:selection}.
The number of signal events is then
determined for each $D$ meson individually
by fitting the invariant mass distributions of the candidate combinations with
a Gaussian to describe the signal and an appropriate background shape. 
The signal mass and width are left free in the fit.
The invariant mass distributions used to extract the 
final number of signal events after all the
vertexing cuts are shown in Fig.~\ref{fig:dall-cst}.
The specific choices for the fits are the following:


a)  \dcdec : the 3-particle invariant mass distribution $m(K \pi \pi)$
    is fitted using a Gaussian for the signal and a linear function describing
    the background shape (Fig.~\ref{fig:dall-cst}a). 
   

b) \dndec: the 2-particle invariant mass distribution 
   $m(K \pi)$~(Fig.~\ref{fig:dall-cst}b)
   is fitted using a Gaussian for the signal, an exponentially falling function
   to describe the combinatorial background and a contribution from 
   wrong-charge combinations. 
   The latter contribution arises since 
   with no particle identification applied, 
   each \dndec\ decay also enters the distribution as a 
   $\bar{D^0} \to K^+ \pi^-$ candidate.
   The contribution of this wrong-charge assignment
   is modelled by a broad Gaussian with
   the mean position, the width and the normalisation relative to the
   correct-charge Gaussian determined from the simulation
   and subsequently kept fixed in the fit.


c) \dsdec: because the $\Dsubs$ decays via the intermediate
  vector meson resonance  $\phi$, 
  it has a restricted 3-body phase space, allowing
  criteria to be applied to suppress the combinatorial background.
  In particular, the 2-particle combination
  $m(K^+ K^-)$ is required to lie within a $\pm 2 \sigma$ window
  ($\pm$11~MeV) around the nominal
  $\phi$ mass. 
  The distribution in the cosine of the angle
  $\theta^*$ (helicity angle), defined as the angle between the $K$-momentum 
  vector and the $\Dsubs$ flight direction, transformed into the 
  $\phi$ rest frame,
  follows  a  $\cos ^2 \theta^*$ shape.  
  A value of $| \cos \theta^* | > 0.4$ is required.  
  The 3-particle invariant mass distribution   $m(K K \pi)$,
  fitted using a Gaussian for the signal and an exponential background function,
  is  shown in Fig.~\ref{fig:dall-cst}c.


d) \dstdec : the $\Delta m$-tagging technique 
is applied \cite{Feldman:1977ir}.
 For all candidate combinations lying within a 3 sigma 
 window ($\pm$ 3.6 MeV)
 of the nominal value of $\Delta m = m(K \pi \pi_s) - m(K \pi)$, 
 the invariant mass distribution
 $m(K \pi)$  is fitted using a Gaussian for the signal and an
 exponential function describing the background shape. The 
 Fig. \ref{fig:dall-cst}d shows this $m(K \pi)$ distribution after
 the cut on  $\Delta m$.


The fitted $D$ meson mass values are all found to be compatible
within errors with the world average values\cite{PDG2000},
and the widths are in agreement with the expected detector resolutions.
For the $D$ meson differential distributions
the data
sample is divided into bins and the number of signal events is extracted
in each bin separately. 
The position and width of the Gaussian describing the signal
are fixed to the values
found in the inclusive sample.
If left free both positions and widths are found to be stable.


\subsection{Acceptance and Efficiency Determination}
\label{simulation}
The Monte Carlo simulation of charm and beauty production
is used to determine the detector acceptance
and the efficiency of the reconstruction and selection cuts.
%
The detector efficiencies are determined to be around $60\%$
for events within the combined geometric and kinematic
acceptances, which vary between $36\%$ and $62\%$
for the different mesons.
%
The lifetime tagging efficiencies are found to be 
19$\%$, 11$\%$, 21$\%$ and 39$\%$, 
for $\Dplus$, $\Dzero$, $\Dsubsplus$ and $\Dstarplus$, respectively.
Migrations due to the limited detector resolution are small for the
chosen bin sizes in the differential distributions
and are corrected by means of Monte Carlo simulation. 


\subsection{Cross Section Determination and Systematic Errors}

The visible cross section $\sigma_{\it vis}$ is defined 
for the  sum of the observed number of  $D$ mesons, $N_D$, for
both particle and antiparticle meson states according to the formula

\begin{equation}
\sigma_{vis}(ep \rightarrow e' D X) =
\frac{N_{D} + N_{\bar{D}}}
  { {\cal L} \cdot  \epsilon \cdot (1 + \delta_{rad}) }\qquad , 
\label{eq:sigman}
\end{equation}
where  $\cal{L}$ denotes the integrated luminosity and $\epsilon$ the total
efficiency including all acceptances and branching ratios, which 
are taken from \cite{PDG2000}.
The radiative corrections $\delta_{rad}$ correct the measured cross
sections to the Born level and are calculated using the
program HECTOR\cite{Arbuzov:1995id}.
They amount to $\delta_{rad} = 2.6 \%$ 
on average and vary between $-4\%$ and $+9\%$ over the kinematic range considered.


The systematic errors on $\sigma_{\it vis}$,
summarised in Table~\ref{tab:syserr},
are dominated by the uncertainty on the
CJC tracking efficiency (conservatively estimated
to $^{+5}_{-1}\% $ per track in this analysis).
A $10 \% $ uncertainty is assigned to the
lifetime cut efficiency to account for a 
possible inaccuracy in the description of the
resolution function in the simulation.
This number is estimated by variations of the
$S_l$-cut and covers the differences between
data and simulation~\cite{Gassner:2002ji}.
The CST efficiency uncertainty is $1\%$ per hit.

The systematic error on the signal extraction is determined 
by variation of the background shapes in the fits.
Backgrounds to the signals from other charm decays
are estimated from Monte Carlo simulations to be at most $3\%$
and are also included in the systematic errors.
The dependence of the simulated acceptances and efficiencies on parameter 
choices made for the simulation (charm mass, parton density distributions,
fragmentation parameters and QCD scales, cf. section~\ref{sec:th}) 
is found to be
less than $\pm 2\% $ in all cases and is included in the 
systematic error.

The uncertainty on the absolute SpaCal electromagnetic energy scale 
contributes a maximal change in $\sigma_{\it vis}$ of $^{+4}_{-9}\%$.
The error due to initial state radiation (ISR) corrections is
estimated to be $\pm 2.6\% $, the average size of the correction itself.
The results obtained when reconstructing the
kinematic variables using the scattered positron have been checked
with the $\Sigma$-method, which combines the hadronic final state and the
positron measurement\cite{Bassler:1994uq}
and are found to be in good agreement. 

The electron detection, the tracking and the vertexing errors 
are to a large extent common between the different mesons
and result in correlated systematic errors of about 15\%.
The $\Dsubs$ meson suffers from a particularly large ($24.7\%$)
uncertainty due to the poorly known branching fraction to $\phi \pi$.

\subsection{Theoretical predictions and uncertainties}
\label{sec:th}
The measured cross sections are compared with LO QCD 
predictions based on the AROMA 2.2~\cite{Ingelman:1996mv}
program, which incorporates the universality ansatz.
The predictions include the contributions from decays
of excited charmed hadrons.
For the nominal predictions the program is run 
using the GRV-98 proton structure function
parametrisation\cite{Gluck:1998xa},
a charm quark mass of $m_c=1.5$ GeV and 
Peterson fragmentation with $\epsilon_c = 0.078$.
The contributions of decays of beauty hadrons is 
simulated using a quark mass of $m_b=4.75$ GeV and
$\epsilon_b = 0.0069$.
It is scaled 
to reproduce the result of \cite{h1-bxsec-dis}.
The renormalisation and factorisation scales are both
set to $\mu = \sqrt{Q^2+ m_t^2(q) + m_t^2({\bar q})}$,
where $m_t(q) = \sqrt{p_t^2 + m_q^2}$ is the 
``transverse'' mass of the quark $q$.
The prediction uncertainties are estimated by simultanously
changing the renormalisation and factorisation scales
by a factor of two or 0.5,
varying $m_c$ between 1.4 GeV and 1.6 GeV, 
and  changing $\epsilon_c$ to $0.048$.
The contributions are added in quadrature. They are 
dominated by the quark mass and fragmentation parameter 
variations (for details see  \cite{Gassner:2002ji}).

%
For the $\Dstar$ meson production cross section
the AROMA prediction is also compared 
with a next to leading order (NLO) calculation 
in the DGLAP scheme \cite{ref:DGLAP}
using the HVQDIS program \cite{Harris:1997zq}.
The GRV98-HO parton densities of the proton\cite{Gluck:1998xa} are used,
supplemented by a Peterson fragmentation 
parametrisation with $\epsilon_c = 0.035$ 
and $\epsilon_b = 0.0033$ as in \cite{Nason:1999zj,Nason:1998ug}.
The renormalisation and factorisation scales are both
set to $\mu = \sqrt{Q^2+ 4 m_q^2}$ 
with the same quark mass values
as are used for the AROMA simulation.
%
The uncertainties on the NLO prediction are estimated 
analogously to the LO case. They are dominated by 
variations of the renormalisation and factorisation 
scales and by
changing the mass $m_c$ between 1.3 GeV and 
1.7 GeV ~\cite{Frixione:nn}.
%
Added in quadrature, the total uncertainties 
are of approximately $15 \% $.
%

\section{Results}

%

\subsection{Differential Production Cross Sections}

\begin{figure}[tb] 
  \begin{center}
    \epsfig{file=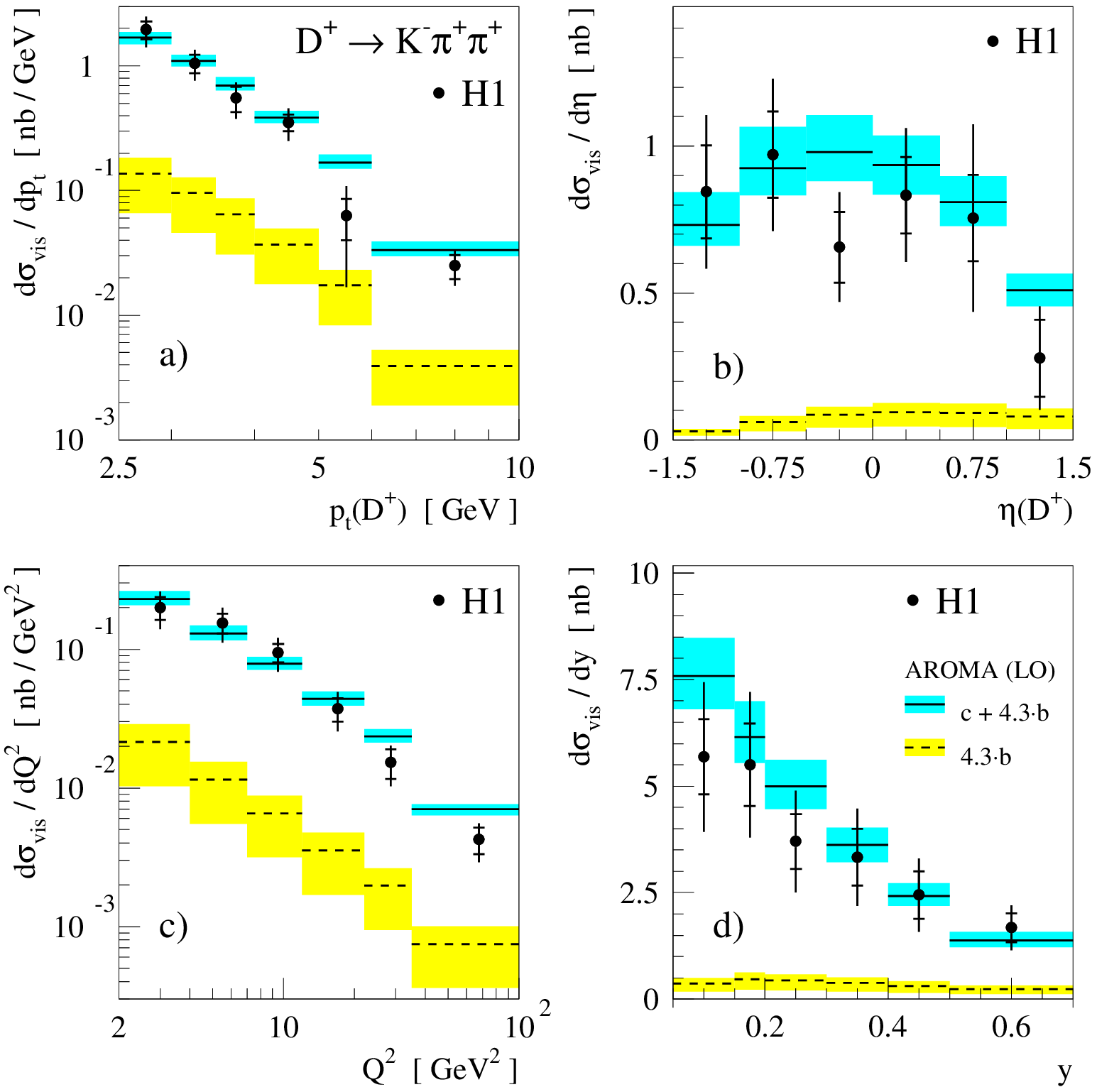,width=16cm}
  \end{center}
  \icaption{ 
   Differential production cross section for \dc \ 
   as a function of 
   a) the \dc \ transverse momentum $p_t(D^+)$, 
   b) the pseudorapidity $\eta(D^+)$, 
   c) the photon virtuality $Q^2$, and d) 
   the inelasticity $y$.
   The symbols denote the values averaged over the bins and 
   are plotted at the bin centres.
   The error bars show the statistical (inner bars) and the  
   total (outer bars) errors, respectively.
   The solid lines show the LO AROMA predictions including scaled 
   beauty contributions, shown separately as dashed lines.
   The shaded bands indicate the uncertainties on the predictions
   (see text).}
  \label{fig:dc-xsec}
\end{figure}

Differential production cross sections are determined  for 
all four $D$ mesons  individually and are tabulated
in Tables~\ref{tab:dplus}-\ref{tab:dstar}.
As an example the results for the \dc\ meson are 
shown  in Fig.~\ref{fig:dc-xsec} as a function 
of the kinematic variables $p_t(D^+)$, $\eta(D^+)$, $Q^2$ and $y$.
For all mesons the cross sections fall rapidly with $Q^2$ and $p_{t}(D)$,
as expected from the hard scattering.
The distributions are well reproduced by the LO AROMA simulation, which is
overlaid in Fig.~\ref{fig:dc-xsec}.
The estimated beauty contributions are shown separately as  dashed lines.
The shaded bands indicate the uncertainties on the predictions.
The differential distributions for the remaining $D^0, D_S$ and $D^*$ mesons 
are equally well described by the LO QCD simulation~\cite{Gassner:2002ji}.

\begin{figure}[htb] 
  \begin{center}
    \epsfig{file=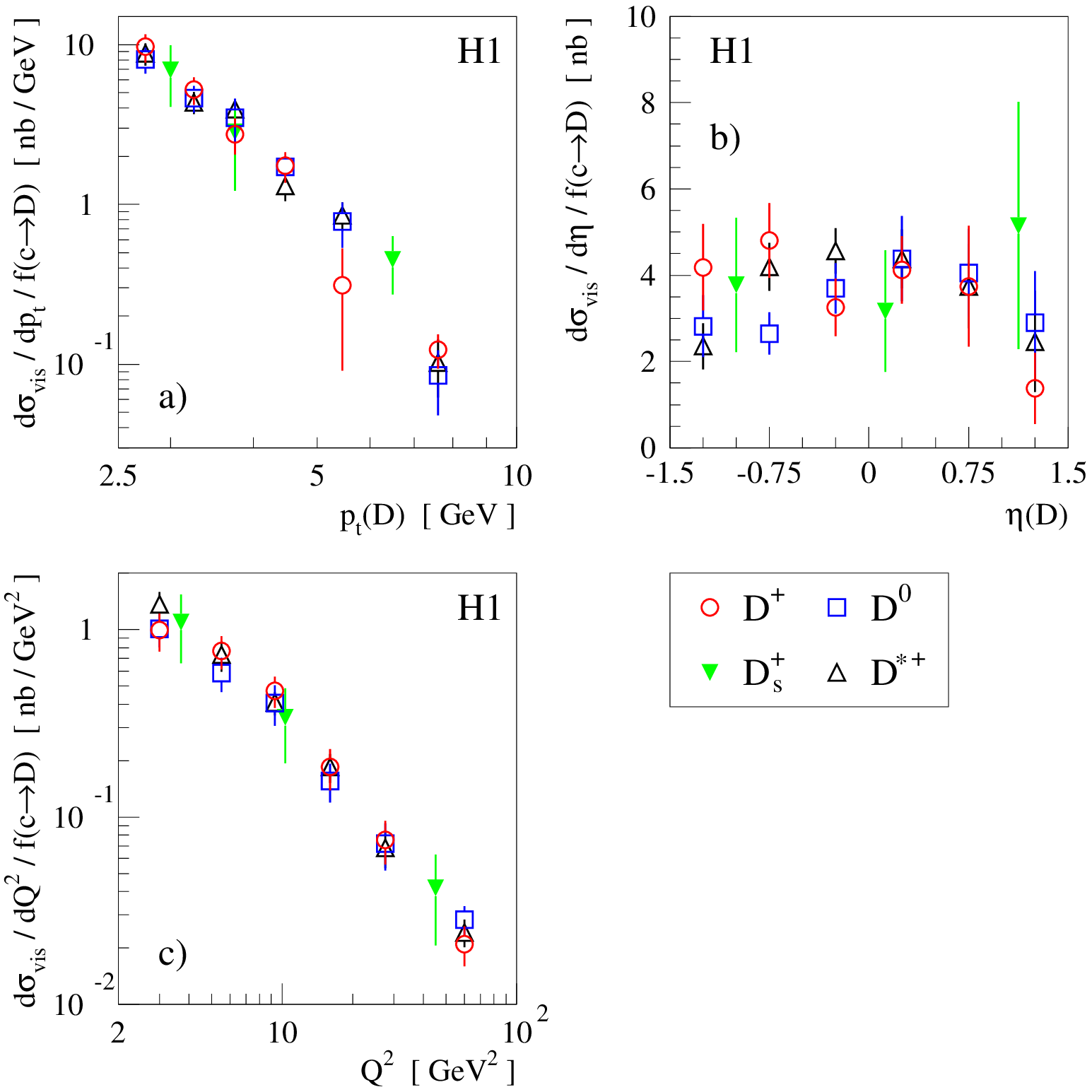}
  \end{center}
  \icaption{ 
  Visible differential production cross sections 
  for all four $D$ mesons,
  divided by  
  their respective measured fragmentation factors,
   are shown  as a function of 
  a) the $D$ meson transverse momentum $p_t(D)$, 
  b) the pseudorapidity $\eta(D)$,  and c) photon virtuality $Q^2$.  
  The error bars
  indicate the quadratic sum of the statistical errors and the 
  systematic errors which are not common to the different
  $D$ mesons.
  An overall common systematic error of $15\%$
  is not shown.
  The cross sections are displayed on the horizontal axis
  at the position that corresponds to the average differential cross section,
  obtained using the AROMA simulation.
  }
  \label{fig:all-xsec}
\end{figure}

For a direct comparison of the differential cross sections extracted for the
four different $D$ mesons, the differential cross sections have been
scaled with the inverse fragmentation factors
derived further below (cf. Table~\ref{tab:ffrag})
and are shown in Fig.~\ref{fig:all-xsec} 
as a function of $p_t(D)$,  $\eta(D)$ and $Q^2$.
Within the present experimental accuracy the normalised cross sections
are compatible with one another.


\subsection{Inclusive Production Cross Sections}

The inclusive electroproduction cross sections are determined
in the visible  kinematic range
of  $2 \le Q^2 \le 100~\gev^2$,
\  $0.05 \le y \le 0.7$,
\ $p_t(D) \ge 2.5~\GeV$ and \ $|\eta(D)| \le 1.5$.
The measured values  for the four $D$ mesons
are summarised in Table~\ref{tab:xsection}.

The $\Dstar$ meson results are consistent
with earlier measurements at
HERA\cite{Adloff:1998vb, Adloff:2001wr, Adloff:2001zj, 
Breitweg:1999ad, Chekanov:2003rb}
once the differences in kinematic regions are taken into account.
For the  $\Dstar$  meson
the visible production cross section has also
been calculated in the next to leading order (NLO) DGLAP
scheme\cite{ref:DGLAP} with the 
HVQDIS program \cite{Harris:1997zq}
(see section~\ref{sec:th}).
The NLO prediction yields a value of $\sigma_{\it vis} = 2.76\pm 0.41$~nb 
for the $\Dstar$,
which is in good agreement with the LO AROMA value of $2.61\pm 0.31$~nb.
\subsection{Fragmentation Factors and Ratios}

In the absence of decays of excited intermediate charm states,
the relative abundance of the various charmed hadrons 
is given by the primary production rate, which is 
directly proportional to $n_D$, the normalisation of the fragmentation
functions (Eq.~\ref{eq:f3}). 
However, the experimentally determined fragmentation factors 
$f^{}(c\rightarrow D)$ 
include all possible decay chains
that result in that particular  $D$ meson,
in addition to the direct production.
Thus, the measured 
pseudoscalar  $D^+$, $D^0$ and $D_s$ mesons
contain a large fraction of mesons produced in $D^*_{(s)}$
decays.

The fragmentation factors $f^{}(c\rightarrow D)$ are deduced from a 
comparison of the measured cross sections $\sigma_{vis}$
with the LO expectation $\sigma_{vis}^{MC}$
after correcting for the beauty contribution
to which an uncertainty of 50\% is attributed.
Explicitly, the following relationship is used:

\begin{eqnarray}
 \displaystyle f^{}(c\rightarrow D) = 
 \frac{\displaystyle \sigma_{\it vis}(ep \rightarrow e'DX) - 
  \sigma^{\it MC}_{\it vis} (ep \rightarrow e' b\bar{b}X' \rightarrow e'DX)}
{  \displaystyle \sigma^{\it MC}_{\it vis} (ep \rightarrow c\bar{c} \rightarrow e'DX)} 
        \cdot \displaystyle f_{\rm wa}(c \rightarrow D).
\end{eqnarray}
The resulting values for $f(c\rightarrow D)$ are listed in Table~\ref{tab:ffrag}.
For comparison, the world average values 
$f_{\rm wa}(c\rightarrow D)$ are also shown. 
The measurements $f(c\rightarrow D)$ agree well within errors with
the world average values $f_{\rm wa}(c\rightarrow D)$. 

Constraints can be explicitly imposed on the measurements,
which improve the experimental accuracy.
The constraint 
\begin{eqnarray}
1       & =  & f(c \to D^+)  + f(c \to {D^0}) + f(c \to D_s)  
  + f(c \to \Lambda_c, \Xi_c, \Omega_c)  \label{eq:constraintsone} 
\end{eqnarray}
introduces contributions of fragmentation to charmed baryons,
which are not determined in this analysis.
World average values~\cite{PDG2000} are taken instead,
yielding a 
total of $f(c \to \Lambda_c, \Xi_c, \Omega_c) = 8.64 \pm 2.50\% $.

The constraint
\begin{eqnarray}
f(c \to D^+)  & =  & f(c \to {D^0}) - 2 \cdot f(c \to D^{*+}) \cdot \BR(D^{*+} \to D^0 \pi) 
\label{eq:constraintstwo}
\end{eqnarray}
is derived by assuming isospin symmetry independently
in both the vector and  pseudoscalar sectors 
($n_{D^{*+}} = n_{D^{*0}}$\ and $n_{D^{+}} = n_{D^{0}}$).
The branching fraction $\BR(D^{*+} \to D^0 \pi)$ is taken as 
$67.6 \pm 0.5~\%$~\cite{PDG2000}.
The $D^{*0}$ mesons, which are not explicitly reconstructed in this analysis,
are assumed to decay entirely into final states containing a $D^0$ meson~\cite{PDG2000}.

The results of a fit for the fragmentation factors using
the constraints of Eq.~\ref{eq:constraintsone}   
and Eq.~\ref{eq:constraintstwo} 
are listed separately in Table~\ref{tab:ffrag} as
$f_{\rm constraint}(c \to D)$.
The fit takes the correlations of the uncertainties
between the different mesons into account.
In comparison with the unconstrained results the errors are reduced and the value
$f(c\to D^0)$ moves closer to the world average, 
whereas the other factors do not change significantly.


Ratios of fragmentation factors
are useful to characterise distinct aspects of fragmentation,
such as the proportions of light quark flavours $u,d$ and $s$
created in the fragmentation process
and the formation of different angular momentum states.
Such quantities  have been measured in light quark fragmentation and
charm fragmentation in $e^+e^-$ 
annihilation~\cite{Nason:1999zj,Albrecht:1991ss,Akers:1994jc,Buskulic:1995gp}.
It is important to verify their applicability for deep inelastic
scattering.

Various ratios are extracted from the $D$ meson fragmentation
factors obtained without the constraints of 
Eqs.~\ref{eq:constraintsone} and ~\ref{eq:constraintstwo}.
The common experimental 
uncertainties largely cancel in this procedure. 
Expressed both in terms of 
the normalisation factors for primary $D$ mesons produced, $n_D$,
and in terms of the  experimentally accessible fragmentation factors 
$ f(c \rightarrow D)$, the ratios considered are: 

\setlength{\jot}{1.5mm}
\begin{eqnarray}
P_V^{d}  = & \frac{n_{D^{*+}}} {n_{D^{*+}}+n_{D^+}}
  & = \frac{f(c \rightarrow D^{*+})}
{f(c\rightarrow D^+) +f(c\rightarrow D^{*+})\cdot {\it \BR}(D^{*+}\rightarrow D^0\pi^+)};
\label{eq:pvd} \\
P_V^{u+d}  = &  \frac{2 \cdot n_{D^{*+}}}{n_{D^{*+}}+n_{D^+}+n_{D^{*0}}+n_{D^0}}
  & = \qquad \qquad \frac{2\cdot f(c \rightarrow D^{*+})}{f(c\rightarrow D^+) +f(c\rightarrow D^0)};
\label{eq:pvud}\\
R_{u/d}  = & \frac{n_{D^{*0}}+n_{D^0}} {n_{D^{*+}}+n_{D^+}}
  & = \frac{ f(c\rightarrow D^0) -f(c\rightarrow D^{*+})\cdot {\it \BR}(D^{*+}\rightarrow D^0\pi^+)}
           { f(c\rightarrow D^+) +f(c\rightarrow D^{*+})\cdot {\it \BR}(D^{*+}\rightarrow D^0\pi^+)};
\label{eq:rud}\\
\gamma_s  = &  \frac{2 \cdot (n_{D^{*+}_{s}}+n_{D^+_{s}})}{n_{D^{*+}}+n_{D^+}+n_{D^{*0}}+n_{D^0}}
  & = \qquad \qquad \frac{ 2\cdot f(c \rightarrow D_s^+)}{ f(c\rightarrow D^+) +f(c\rightarrow D^0)}.
\label{eq:gammas}
\end{eqnarray}


\noindent
The fragmentation characteristics derived according to Eqs.~\ref{eq:pvd}-\ref{eq:gammas}
are summarised in Table~\ref{tab:fratios}.
The ratio $P_{V}$ measures the fraction of vector mesons produced in the fragmentation chain.
This ratio has been evaluated separately for the $c\bar{d}$-quark combination
(Eq.~\ref{eq:pvd}) 
and the sum of $c\bar{u}$ and $c\bar{d}$ quark combinations (Eq.~\ref{eq:pvud}).
Isospin invariance $ f(c\rightarrow D^{*+}) = f(c\rightarrow D^{*0}) $
has been assumed for $P_V^{u+d}$
and the result agrees well with values
measured at LEP~\cite{ff:aleph,ff:delphi}.
The $P_V^d$ result is larger than the LEP value by around 1.5 sigma.

The ratio $R_{u/d}$ of probabilities for a charm quark to hadronise together
with a $u$ or $d$-quark
agrees well with a value of unity, as is also the case
at LEP~\cite{ff:aleph}.
The suppression of $s$-quarks with respect to $u$ and $d$ quarks
is measured by the factor $\gamma_{s}$.
The value of $\gamma_{s}$ extracted in this analysis suggests a
suppression of almost a factor 3.
This observation agrees well with the suppression factor measured in
the fragmentation of primary light quarks and with measurements of charm
fragmentation  in $e^+e^-$~\cite{ff:ado}.


\section{Conclusions}

Production cross sections of the charmed mesons
$\Dplus$, $\Dzero$, $\Dsubs$ and $D^{*+}$ 
are measured in deep inelastic
$ep$ collisions at HERA. 
The measurements rely on the precise reconstruction of the
$D$ meson production and decay vertices using the
central silicon tracker of the H1 detector. 
The differential production cross sections of all
four $D$ mesons measured in the same kinematic region
show a very similar dependence on 
the transverse momentum $p_t(D)$ and the pseudorapidity $\eta(D)$ of the 
charmed hadron, as well as on the photon virtuality $Q^2$.

Based on the measured cross sections, the 
fragmentation factors $ f(c\rightarrow D)$ are extracted. 
They are found to agree with the world average values, which are
dominated by measurements made in $e^+e^-$ collisions. 
Based on ratios of these $ f(c\rightarrow D)$ the fragmentation-sensitive
parameters  $P_V$, $R_{u/d}$ and $\gamma_{s}$ are determined. 
They are also in agreement with the world average values.

The measurements of the cross sections and of the
fragmentation-sensitive ratios thus
support the validity of the  commonly made assumptions 
that the charm quark production and the fragmentation processes 
factorise and that the non-perturbative component 
of the hadronisation $D_D^{(c)}(z)$
is well described by a universal fragmentation ansatz,
independent of the hard scattering process and
of the charm production scale.

\section*{Acknowledgements}

We are grateful to the HERA machine group whose outstanding
efforts have made this experiment possible. 
We thank
the engineers and technicians for their work in constructing and
maintaining the H1 detector, our funding agencies for 
financial support, the
DESY technical staff for continual assistance
and the DESY directorate for support and for the
hospitality which they extend to the non DESY 
members of the collaboration.

\clearpage 
\newpage


\clearpage

%

\vspace{2cm}



\clearpage

\begin{table}[htb]
\renewcommand{\arraystretch}{1.1}
\begin{center}
 \begin{tabular}{|c|c|c|c|c|}
  \hline
  {\bf Selection criteria} & \dc & \dn & \ds & \dst \\
  \hline
  min $p_t(K)\ [ \mev ]\ $     & 500 & 800 & 400 & 250 \\
  min $p_t(\pi)\ \ [ \mev ]\ $ & 400 & 800 & 400 & 250 \\
  min $p_t(\pi_s)\ [ \mev ]\ $ &   - &   - &   - & 140 \\
  \hline
   min impact par. sig. $ S_d $ & 2.5 & 2 &   1 &  -1 \\
  min decay length sig. $ S_l $       &   5 &   3 &   2 &   1 \\
  \hline
  max decay length error $ \sigma_l $              & \multicolumn{4}{c|}{300~$\mu$m} \\
  min fit probability ${\cal P}_{\it VF} $          & \multicolumn{4}{c|}{ 0.05} \\
  \hline
 \end{tabular}
\end{center}
\icaption{  
  Selection criteria for each of the four $D$ mesons.
  Listed are the minimal transverse momenta of the daughter tracks
  (cuts on the slow pion $\pi_s$ are only applicable for the $D^*$ meson) 
  and the requirements on the vertexing parameters (see text). }
\label{tab:selection}
\end{table}
%

\begin{table}[htb]
 \renewcommand{\arraystretch}{1.4}
 \centering
 \begin{tabular}{|c|c|c|c|c|}
  \hline
  {\bf Source of uncertainty} & \dc & \dn & \ds & \dst \\
\hline
CJC efficiency  & $^{+15.0}_{\ -3.0}$  & $^{+10.0}_{\ -2.0}$  & $^{+15.0}_{\ -3.0}$  & $^{+15.0}_{\ -3.0}$  \\
\hline
CST efficiency  & $\pm 5.6$  &  $\pm 3.6$  & $\pm 5.4$  & $\pm 3.6$  \\
\hline
Lifetime tag & \multicolumn{4}{c|}{$\pm 10.0$}  \\
\hline
Signal extraction & $^{+1.7}_{-0.4}$  & $^{\ +4.9}_{-13.4}$  & $^{\ +1.4}_{-12.1}$  & $^{+3.7}_{-3.2}$  \\
\hline
Branching fraction & $\pm6.7$ & $\pm2.3$ & $\pm24.7$ \ & $\pm2.3$ \\
\hline
\hline
Total systematic error & $^{+21.2}_{-16.0}$ & $^{+16.5}_{-19.9}$ & $^{+32.0}_{-31.8}$ & $^{+20.0}_{-15.2}$ \\

\hline
 \end{tabular}
\icaption{ 
  Percentage systematic errors on the inclusive cross sections
  for the different $D$ mesons. Only the dominant
  contributions and the total error are shown. }
\label{tab:syserr}
\end{table}
%

%
%
\begin{table}[htb]
 \renewcommand{\arraystretch}{1.2}
\begin{center}
 \begin{tabular}{|c|c|c|c|c|}
  \hline
  {\bf Cross section [nb]} & \dc & \dn & \ds & \dst \\
  \hline 
$\sigma_{\it vis}(ep \rightarrow eDX)$   &
  $2.16$      & $6.53$     & $1.67$ & $2.90$   \\
Stat.\ error          &
$\pm 0.19$ & $\pm 0.49$ & $\pm 0.41$ & $\pm 0.20$  \\
Syst.\ error           &
$^{+0.46}_{-0.35}$ & $^{+1.06}_{-1.30}$ & $^{+0.54}_{-0.54}$ & $^{+0.58}_{-0.44}$ \\
  \hline  
AROMA LO prediction $ \sigma_{\it vis}$ & $2.45$ & $5.54$ & $1.15$ & $2.61$   \\
   Prediction uncertainty          &
$\pm 0.30$ & $\pm 0.69$ &  $\pm 0.30$&  $\pm 0.31 $  \\
  \hline
  Estimated beauty contribution & $10\%$ & $9\%$ & $17\%$ &  $7\%$ \\
  \hline
\end{tabular}
\end{center}
   \icaption{ 
  Inclusive charmed meson electroproduction cross sections for the four 
  meson states in the visible kinematic range, defined by 
  $2\, \le \,Q^2 \le 100\, \GeV^2,\
\ 0.05 \le y  \le  0.7, \ p_t(D) \ge 2.5\,\GeV$ and $|\eta(D)|  \le  1.5. $   
  Also given are the predictions for $D$ meson production (including
  the beauty contribution) based on a LO Monte Carlo simulation. }
\label{tab:xsection}
\end{table}

%
\begin{table}[tbh]
\renewcommand{\arraystretch}{1.35}
\begin{center}
 \begin{tabular}{|c|c|c|c|c|}
  \hline
  {\bf Fragmentation factors} & \dc & \dn & \ds & \dst \\
  \hline 
  $ f(c \to D)$         & $0.202$ &  $0.658$  & $0.156$ & $0.263$ \\
 Stat.\ error            &  $\pm 0.020$ &  $\pm 0.054$ &  $\pm 0.043$ &  $\pm 0.019 $ \\
 Syst.\ error            & $^{+0.045}_{-0.033}$ & $^{+0.117}_{-0.142}$ & $^{+0.036}_{-0.035}$ & $^{+0.056}_{-0.042}$\\
 Theo.\ error            & $^{+0.029}_{-0.021}$ & $^{+0.086}_{-0.048}$ & $^{+0.050}_{-0.046}$ & $^{+0.031}_{-0.022}$ \\
\hline
$f_{\rm constraint}(c \to D)$ & $0.203$ &  $0.560$  & $0.151$ & $0.263$  \\
 Total error                & $\pm 0.026$ &  $\pm 0.046$ &  $\pm 0.055$ &  $\pm 0.032 $ \\
\hline
 $f_{\rm wa}(c \to D)$ & $0.232 \pm 0.018$ & $0.549 \pm 0.026$ & $0.101  \pm 0.027$  & $0.235  \pm 0.010$  \\
  \hline
 \end{tabular}
\end{center}
   \icaption{ 
  Fragmentation factors deduced from the measured cross sections.
  The small $b$~contributions are subtracted. 
  Also listed is the result of the
  constrained fit $f_{\rm constraint}(c \to D)$, its error (see text), and
  the world average numbers, given as $f_{\rm wa}(c \to D)$. 
  The theoretical errors include the branching ratio uncertainty
  and the model dependences of the acceptance determination.
  }
\label{tab:ffrag}
\end{table}

%
\begin{table}[tbh]
\renewcommand{\arraystretch}{1.1}
\begin{center}
 \begin{tabular}{|c|cccc|ccr|}
   \hline
  {\bf Ratio} & \multicolumn{4}{c|}{{\bf this measurement}} 
                   & \multicolumn{3}{c|}{{\bf $e^+e^-$ experiments}} \\
     & value   & stat.error & syst.error & theo.error & value & error & ref. \\
\hline

$P_V^d$     & $0.693$ & $\pm$  $0.045$ & $\pm$  $0.004$ & $\pm$ $0.009$ 
            & $0.595$ & $\pm$  $0.045$ & \cite{ff:aleph} \\

$P_V^{u+d}$ & $0.613$ & $\pm$  $0.061$ & $\pm$  $0.033$ & $\pm$ $0.008$
            & $0.620$ & $\pm$  $0.014$ & \cite{ff:delphi}\\

$R_{u/d}$   & $1.26$  & $\pm$  $0.20$  & $\pm$  $0.11$  & $\pm$ $0.04$ 
            & $1.02$  & $\pm$  $0.12$  & \cite{ff:aleph}  \\

$\gamma_s$  & $0.36$  & $\pm$  $0.10$  & $\pm$  $0.01$  & $\pm$ $0.08$ 
            & $0.31$  & $\pm$  $0.07$  & \cite{ff:ado}   \\
\hline
 \end{tabular}
\end{center}
   \icaption{Fragmentation characteristics derived according to 
   Eqs.~\ref{eq:pvd}-\ref{eq:gammas} from the ratios
   of the fragmentation factors of Table~\ref{tab:ffrag}. 
   Values measured by $e^+e^-$  experiments are also quoted.
   The theoretical errors include the branching ratio uncertainty
   and the model dependences of the acceptance determination.
}
\label{tab:fratios}
\end{table}

\clearpage


%
%
%
\begin{center}
\begin{table}[h]
 \vspace*{1cm}
 \hspace*{1.5cm}
 \renewcommand{\arraystretch}{1.6}
   \scriptsize
%
   \begin{tabular}[t]{|@{\quad $[$\ }c@{\ $]$\quad }|c|c|c|c|c|c|}
    \hline
    \multicolumn{1}{|c|}{\small $p_t(D^+)$} & \small ${\rm d}\sigma_{vis}/{\rm d}p_t$ &
    \multicolumn{5}{c|}{errors \ $[\unit{nb/GeV}\ ]$} \\
    \cline{3-7}
    \multicolumn{1}{|c|}{$[\unit{GeV}\:]$} & $[\unit{nb/GeV}\ ]$ &
    statistical & \multicolumn{2}{c|}{experimental} & \multicolumn{2}{c|}{theoretical} \\
    \hline
    2.5, \,\ 3.0  & \small 1.95\; & $\pm0.32\; $ & $+0.41\;$ & $-0.34\; $ & $+0.13\;$ & $-0.14\;$ \\
    3.0, \,\ 3.5  & \small 1.05\; & $\pm0.18\; $ & $+0.22\;$ & $-0.17\; $ & $+0.07\;$ & $-0.08\;$ \\
    3.5, \,\ 4.0  & \small 0.556 & $\pm0.128 $ & $+0.121$ & $-0.090$ & $+0.039$ & $-0.039$ \\
    4.0, \,\ 5.0  & \small 0.353 & $\pm0.054$ & $+0.104$ & $-0.054$ & $+0.026$ & $-0.024$ \\
    5.0, \,\ 6.0  & \small 0.063 & $\pm0.023$ & $+0.072$ & $-0.010$ & $+0.004$ & $-0.005$ \\
    6.0,    10.0  & \small 0.025 & $\pm0.005$ & $+0.005$ & $-0.004$ & $+0.002$ & $-0.002$ \\
    \hline
    \multicolumn{7}{c}{}\\
    \hline
    \multicolumn{1}{|c|}{\small $\eta(D^+)$} & \small ${\rm d}\sigma_{vis}/{\rm d}\eta$ &
    \multicolumn{5}{c|}{errors \ $[\unit{nb}\ ]$} \\
    \cline{3-7}
    \multicolumn{1}{|c|}{} & $[\unit{nb}\ ]$ &
    statistical & \multicolumn{2}{c|}{experimental} & \multicolumn{2}{c|}{theoretical} \\
    \hline
     -1.50,  -1.00  & \small 0.84 & $\pm0.16$ & $+0.25$ & $-0.12$ & $+0.06$ & $-0.08$ \\
     -1.00,  -0.50  & \small 0.97 & $\pm0.15$ & $+0.20$ & $-0.14$ & $+0.07$ & $-0.07$ \\
     -0.50, \ 0.00  & \small 0.66 & $\pm0.12$ & $+0.14$ & $-0.10$ & $+0.05$ & $-0.04$ \\
    \ 0.00, \ 0.50  & \small 0.83 & $\pm0.13$ & $+0.18$ & $-0.13$ & $+0.06$ & $-0.06$ \\
    \ 0.50, \ 1.00  & \small 0.76 & $\pm0.15$ & $+0.36$ & $-0.22$ & $+0.06$ & $-0.05$ \\
    \ 1.00, \ 1.50  & \small 0.28 & $\pm0.13$ & $+0.15$ & $-0.09$ & $+0.02$ & $-0.03$ \\
    \hline
    \multicolumn{7}{c}{\parbox[l][8mm][c]{11cm}{}} \\
    \hline
    \multicolumn{1}{|c|}{\small $Q^2$} & \small ${\rm d}\sigma_{vis}/{\rm d}Q^2$ &
    \multicolumn{5}{c|}{errors \ $[\unit{nb/GeV}^2\ ]$} \\
    \cline{3-7}
    \multicolumn{1}{|c|}{$[\unit{GeV}^2\,]$} & $[\unit{nb/GeV}^2\ ]$ &
    statistical & \multicolumn{2}{c|}{experimental} & \multicolumn{2}{c|}{theoretical} \\
    \hline
    \,\ 2, \ \ \ 4  & \small 0.200\; & $\pm0.038\; $ & $+0.047\; $ & $-0.033\; $ & $+0.014\; $ & $-0.015\;$ \\
    \,\ 4, \ \ \ 7  & \small 0.155\; & $\pm0.025\; $ & $+0.034\; $ & $-0.025\; $ & $+0.011\;$ & $-0.011\; $ \\
    \,\ 7,  \,\ 12  & \small 0.095\; & $\pm0.015\; $ & $+0.021\; $ & $-0.014\; $ & $+0.008\;$ & $-0.007\; $ \\
       12,  \,\ 22  & \small 0.0373 & $\pm0.0072$ & $+0.0105$ & $-0.0060$ & $+0.0025$ & $-0.0027$ \\
       22,  \,\ 35  & \small 0.0153 & $\pm0.0036$ & $+0.0033$ & $-0.0026$ & $+0.0010$ & $-0.0012$ \\
       35,     100  & \small 0.0042 & $\pm0.0009$ & $+0.0009$ & $-0.0007$ & $+0.0003$ & $-0.0003$ \\
    \hline
   \end{tabular}

    \icaption{\label{tab.dc.xsec.bins} 
 The bin-averaged  single differential \dc\ production cross section
 measurements $\sigma_{vis}(ep\rightarrow eD^+X)$
with statistical, experimental systematic and 
theoretical errors.
The latter include the uncertainties of the
branching ratio and the model dependence of the acceptance
determination.}
\label{tab:dplus}
\end{table}
\end{center}
\clearpage
%
%
%
\begin{center}
\begin{table}[th]
 \vspace*{1cm}
 \hspace*{1.5cm}
 \renewcommand{\arraystretch}{1.6}
   \scriptsize
%
   \begin{tabular}[t]{|@{\quad $[$\ }c@{\ $]$\quad }|c|c|c|c|c|c|}
    \hline
    \multicolumn{1}{|c|}{\small $p_t(D^0)$} & \small ${\rm d}\sigma_{vis}/{\rm d}p_t$ &
    \multicolumn{5}{c|}{errors \ $[\unit{nb/GeV}\ ]$} \\
    \cline{3-7}
    \multicolumn{1}{|c|}{$[\unit{GeV}\,]$} & $[\unit{nb/GeV}\ ]$ &
    statistical & \multicolumn{2}{c|}{experimental} & \multicolumn{2}{c|}{theoretical} \\
    \hline
    2.5, \,\ 3.0  & \small 5.29\; & $\pm0.72\;$ & $+1.11\;$ & $-0.89\;$ & $+0.13\;$ & $-0.18\;$ \\
    3.0, \,\ 3.5  & \small 3.04\; & $\pm0.41\;$ & $+0.50\;$ & $-0.78\;$ & $+0.13\;$ & $-0.08\;$ \\
    3.5, \,\ 4.0  & \small 2.29\; & $\pm0.34\;$ & $+0.81\;$ & $-0.62\;$ & $+0.09\;$ & $-0.06\;$ \\
    4.0, \,\ 5.0  & \small 1.13\; & $\pm0.15\;$ & $+0.19\;$ & $-0.21\;$ & $+0.03\;$ & $-0.04\;$ \\
    5.0, \,\ 6.0  & \small 0.516 & $\pm0.120$ & $+0.139$ & $-0.129$ & $+0.018$ & $-0.015$ \\
    6.0,    10.0  & \small 0.056 & $\pm0.018$ & $+0.016$ & $-0.021$ & $+0.002$ & $-0.003$ \\
    \hline
    \multicolumn{7}{c}{}\\
    \hline
    \multicolumn{1}{|c|}{\small $\eta(D^0)$} & \small ${\rm d}\sigma_{vis}/{\rm d}\eta$ &
    \multicolumn{5}{c|}{errors \ $[\unit{nb}\ ]$} \\
    \cline{3-7}
    \multicolumn{1}{|c|}{} & $[\unit{nb}\ ]$ &
    statistical & \multicolumn{2}{c|}{experimental} & \multicolumn{2}{c|}{theoretical} \\
    \hline
     -1.50,  -1.00  & \small 1.85 & $\pm0.38$ & $+0.34$ & $-0.46$ & $+0.09$ & $-0.05$ \\
     -1.00,  -0.50  & \small 1.74 & $\pm0.30$ & $+0.28$ & $-0.28$ & $+0.04$ & $-0.07$ \\
     -0.50, \ 0.00  & \small 2.43 & $\pm0.34$ & $+0.42$ & $-0.36$ & $+0.06$ & $-0.11$ \\
    \ 0.00, \ 0.50  & \small 2.88 & $\pm0.39$ & $+0.53$ & $-0.87$ & $+0.10$ & $-0.07$ \\
    \ 0.50, \ 1.00  & \small 2.67 & $\pm0.37$ & $+0.52$ & $-0.87$ & $+0.10$ & $-0.07$ \\
    \ 1.00, \ 1.50  & \small 1.91 & $\pm0.41$ & $+0.70$ & $-0.77$ & $+0.07$ & $-0.08$ \\
    \hline
    \multicolumn{7}{c}{\parbox[l][8mm][c]{11cm}{}} \\
    \hline
    \multicolumn{1}{|c|}{\small $Q^2$} & \small ${\rm d}\sigma_{vis}/{\rm d}Q^2$ &
    \multicolumn{5}{c|}{errors \ $[\unit{nb/GeV}^2\ ]$} \\
    \cline{3-7}
    \multicolumn{1}{|c|}{$[\unit{GeV}^2\,]$} & $[\unit{nb/GeV}^2\ ]$ &
    statistical & \multicolumn{2}{c|}{experimental} & \multicolumn{2}{c|}{theoretical} \\
    \hline
    \,\ 2, \ \ \ 4  & \small 0.662\; & $\pm0.099\;$ & $+0.113\;$ & $-0.161\;$ & $+0.024\;$ & $-0.019\;$ \\
    \,\ 4, \ \ \ 7  & \small 0.386\; & $\pm0.058\;$ & $+0.079\;$ & $-0.088\;$ & $+0.020\;$ & $-0.010\;$ \\
    \,\ 7,  \,\ 12  & \small 0.267\; & $\pm0.043\;$ & $+0.066\;$ & $-0.057\;$ & $+0.008\;$ & $-0.007\;$ \\
       12,  \,\ 22  & \small 0.102\; & $\pm0.018\;$ & $+0.021\;$ & $-0.023\;$ & $+0.003\;$ & $-0.003\;$ \\
       22,  \,\ 35  & \small 0.0474 & $\pm0.0113$ & $+0.0096$ & $-0.0107$ & $+0.0013$ & $-0.0013$ \\
       35,     100  & \small 0.0186 & $\pm0.0027$ & $+0.0032$ & $-0.0038$ & $+0.0005$ & $-0.0007$ \\
    \hline
   \end{tabular}

    \icaption{\label{tab.dn.xsec.bins} 
The bin-averaged single differential \dn\ production cross section
measurements $\sigma_{vis}(ep\rightarrow eD^0X)$
with statistical, experimental systematic and 
theoretical errors.
The latter include the uncertainties of the
branching ratio and the model dependence of the acceptance
determination.}
\label{tab:dzero}
\end{table}
\end{center}
\clearpage
%
%
%
\begin{center}
\begin{table}[th]
 \vspace*{1cm}
 \hspace*{1.5cm}
 \renewcommand{\arraystretch}{1.6}
   \scriptsize
%
   \begin{tabular}[t]{|@{\quad $[$\ }c@{\ $]$\quad }|c|c|c|c|c|c|}
    \hline
    \multicolumn{1}{|c|}{\small $p_t(D_s^+)$} & \small ${\rm d}\sigma_{vis}/{\rm d}p_t$ &
    \multicolumn{5}{c|}{errors \ $[\unit{nb/GeV}\ ]$} \\
    \cline{3-7}
    \multicolumn{1}{|c|}{$[\unit{GeV}\,]$} & $[\unit{nb/GeV}\ ]$ &
    statistical & \multicolumn{2}{c|}{experimental} & \multicolumn{2}{c|}{theoretical} \\
    \hline
    2.5, \,\ 3.5  & \small 1.09\; & $\pm0.33\;$ & $+0.22\;$ & $-0.27\;$ & $+0.27\;$ & $-0.27\;$ \\
    3.5, \,\ 4.0  & \small 0.44\; & $\pm0.20\;$ & $+0.16\;$ & $-0.09\;$ & $+0.11\;$ & $-0.11\;$ \\
    4.0,    10.0  & \small 0.071 & $\pm0.021$ & $+0.016$ & $-0.012$ & $+0.018$ & $-0.018$ \\
    \hline
    \multicolumn{7}{c}{}\\
    \hline
    \multicolumn{1}{|c|}{\small $\eta(D_s^+)$} & \small ${\rm d}\sigma_{vis}/{\rm d}\eta$ &
    \multicolumn{5}{c|}{errors \ $[\unit{nb}\ ]$} \\
    \cline{3-7}
    \multicolumn{1}{|c|}{} & $[\unit{nb}\ ]$ &
    statistical & \multicolumn{2}{c|}{experimental} & \multicolumn{2}{c|}{theoretical} \\
    \hline
     -1.50,  -0.50  & \small 0.59 & $\pm0.19$ & $+0.12$ & $-0.10$ & $+0.15$ & $-0.15$ \\
     -0.50, \ 0.75  & \small 0.50 & $\pm0.17$ & $+0.10$ & $-0.12$ & $+0.12$ & $-0.12$ \\
    \ 0.75, \ 1.50  & \small 0.81 & $\pm0.25$ & $+0.43$ & $-0.29$ & $+0.20$ & $-0.21$ \\
    \hline
    \multicolumn{7}{c}{\parbox[l][8mm][c]{11cm}{}} \\
    \hline
    \multicolumn{1}{|c|}{\small $Q^2$} & \small ${\rm d}\sigma_{vis}/{\rm d}Q^2$ &
    \multicolumn{5}{c|}{errors \ $[\unit{nb/GeV}^2\ ]$} \\
    \cline{3-7}
    \multicolumn{1}{|c|}{$[\unit{GeV}^2\,]$} & $[\unit{nb/GeV}^2\ ]$ &
    statistical & \multicolumn{2}{c|}{experimental} & \multicolumn{2}{c|}{theoretical} \\
    \hline
    \,\ 2, \ \ \ 6  & \small 0.172\; & $\pm0.044\;$ & $+0.048\;$ & $-0.038\;$ & $+0.043\;$ & $-0.043\;$ \\
    \,\ 6,  \,\ 16  & \small 0.053\; & $\pm0.017\;$ & $+0.012\;$ & $-0.012\;$ & $+0.013\;$ & $-0.013\;$ \\
       16,     100  & \small 0.0065 & $\pm0.0028$ & $+0.0014$ & $-0.0013$ & $+0.0016$ & $-0.0016$ \\
    \hline
   \end{tabular}

    \icaption{\label{tab.ds.xsec.bins} 
The bin-averaged single differential \ds\ production cross section
measurements $\sigma_{vis}(ep\rightarrow eD_s^+X)$ 
with statistical, experimental systematic and 
theoretical errors.
The latter include the uncertainties of the
branching ratio and the model dependence of the acceptance
determination.}
\label{tab:dsubs}
\end{table}
\end{center}
\clearpage
%
%
\begin{center}
\begin{table}[ht]
 \hspace*{1.5cm}
 \renewcommand{\arraystretch}{1.6}
   \scriptsize
%
   \begin{tabular}[t]{|@{\quad $[$\ }c@{\ $]$\quad }|c|c|c|c|c|c|}
    \hline
    \multicolumn{1}{|c|}{\small $p_t(D^{*+})$} & \small ${\rm d}\sigma_{vis}/{\rm d}p_t$ &
    \multicolumn{5}{c|}{errors \ $[\unit{nb/GeV}\ ]$} \\
    \cline{3-7}
    \multicolumn{1}{|c|}{$[\unit{GeV}\,]$} & $[\unit{nb/GeV}\ ]$ &
    statistical & \multicolumn{2}{c|}{experimental} & \multicolumn{2}{c|}{theoretical} \\
    \hline
    2.5, \,\ 3.0  & \small 2.31\; & $\pm0.28\;$ & $+0.50\;$ & $-0.40\;$ & $+0.07\;$ & $-0.06\;$ \\
    3.0, \,\ 3.5  & \small 1.15\; & $\pm0.16\;$ & $+0.23\;$ & $-0.19\;$ & $+0.03\;$ & $-0.04\;$ \\
    3.5, \,\ 4.0  & \small 1.03\; & $\pm0.13\;$ & $+0.21\;$ & $-0.16\;$ & $+0.03\;$ & $-0.03\;$ \\
    4.0, \,\ 5.0  & \small 0.344 & $\pm0.055$ & $+0.085$ & $-0.055$ & $+0.011$ & $-0.009$ \\
    5.0, \,\ 6.0  & \small 0.224 & $\pm0.041$ & $+0.048$ & $-0.037$ & $+0.006$ & $-0.009$ \\
    6.0,    10.0  & \small 0.027 & $\pm0.007$ & $+0.015$ & $-0.004$ & $+0.001$ & $-0.001$ \\
    \hline
    \multicolumn{7}{c}{}\\
    \hline
    \multicolumn{1}{|c|}{\small $\eta(D^{*+})$} & \small ${\rm d}\sigma_{vis}/{\rm d}\eta$ &
    \multicolumn{5}{c|}{errors \ $[\unit{nb}\ ]$} \\
    \cline{3-7}
    \multicolumn{1}{|c|}{} & $[\unit{nb}\ ]$ &
    statistical & \multicolumn{2}{c|}{experimental} & \multicolumn{2}{c|}{theoretical} \\
    \hline
     -1.50,  -1.00  & \small 0.62 & $\pm0.13$ & $+0.14$ & $-0.09$ & $+0.02$ & $-0.02$ \\
     -1.00,  -0.50  & \small 1.10 & $\pm0.13$ & $+0.23$ & $-0.14$ & $+0.03$ & $-0.03$ \\
     -0.50, \ 0.00  & \small 1.20 & $\pm0.14$ & $+0.24$ & $-0.15$ & $+0.03$ & $-0.03$ \\
    \ 0.00, \ 0.50  & \small 1.15 & $\pm0.16$ & $+0.22$ & $-0.20$ & $+0.03$ & $-0.04$ \\
    \ 0.50, \ 1.00  & \small 0.99 & $\pm0.15$ & $+0.23$ & $-0.31$ & $+0.03$ & $-0.02$ \\
    \ 1.00, \ 1.50  & \small 0.65 & $\pm0.16$ & $+0.32$ & $-0.25$ & $+0.03$ & $-0.02$ \\
    \hline
    \multicolumn{7}{c}{\parbox[l][8mm][c]{11cm}{}} \\
    \hline
    \multicolumn{1}{|c|}{\small $Q^2$} & \small ${\rm d}\sigma_{vis}/{\rm d}Q^2$ &
    \multicolumn{5}{c|}{errors \ $[\unit{nb/GeV}^2\ ]$} \\
    \cline{3-7}
    \multicolumn{1}{|c|}{$[\unit{GeV}^2\,]$} & $[\unit{nb/GeV}^2\ ]$ &
    statistical & \multicolumn{2}{c|}{experimental} & \multicolumn{2}{c|}{theoretical} \\
    \hline
    \,\ 2, \ \ \ 4  & \small 0.359\; & $\pm0.043\;$ & $+0.079\;$ & $-0.062\;$ & $+0.014\;$ & $-0.009\;$ \\
    \,\ 4, \ \ \ 7  & \small 0.193\; & $\pm0.027\;$ & $+0.045\;$ & $-0.036\;$ & $+0.005\;$ & $-0.006\;$ \\
    \,\ 7,  \,\ 12  & \small 0.107\; & $\pm0.014\;$ & $+0.021\;$ & $-0.015\;$ & $+0.003\;$ & $-0.004\;$ \\
       12,  \,\ 22  & \small 0.0486 & $\pm0.0068$ & $+0.0109$ & $-0.0084$ & $+0.0014$ & $-0.0013$ \\
       22,  \,\ 35  & \small 0.0180 & $\pm0.0038$ & $+0.0036$ & $-0.0026$ & $+0.0005$ & $-0.0007$ \\
       35,     100  & \small 0.0064 & $\pm0.0010$ & $+0.0013$ & $-0.0009$ & $+0.0002$ & $-0.0002$ \\
    \hline
   \end{tabular}

    \icaption{\label{tab.dstar.cst.xsec.bins} 
The bin-averaged  single differential \dst\ production cross section
measurements $\sigma_{vis}(ep\rightarrow eD^{*+}X)$ 
with statistical, experimental systematic and 
theoretical errors.
The latter include the uncertainties of the
branching ratio and the model dependence of the acceptance
determination.
The \dn\ signals used for the \dst\ measurement 
fulfill life time tagging requirements. 
}
\label{tab:dstar}
\end{table}
\end{center}
\clearpage
%

\end{document}